\documentstyle[eqsecnum,pra,aps]{revtex}
\begin{document}
\draft
\newcommand{\const}{\rm const}
%@twocolumnfalse\endcsname

\title{Collective excitations in Bose-Einstein condensates in triaxially
anisotropic parabolic traps}

\author{Andr\'as Csord\'as,}
\address{Research Group for Statistical Physics of the
Hungarian Academy of Sciences,\\
M\'uzeum krt. 6--8, H-1088 Budapest,
Hungary}
\author{Robert Graham,}
\address{Fachbereich Physik,
Universit\"at-Gesamthochschule Essen,\\
45117 Essen,Germany}

\date{\today}
\maketitle

\begin{abstract}
The wave equation of low-frequency density waves in Bose-Einstein condensates
at vanishing temperature in arbitrarily anisotropic harmonic traps is
separable in elliptic coordinates, provided the condensate can be treated in
the Thomas-Fermi approximation. We present a complete solution of the mode
functions, which are polynomials of finite order, and their eigenfrequencies
which are characterized by three integer quantum numbers.
\end{abstract}


%\vskip2pc]
%\narrowtext

%%%%%%%%%%%%%%%%%%%%%%%%%%%%%%%%%%%%%%%%%%%%%%%%%%%%%
\section{Introduction}
%%%%%%%%%%%%%%%%%%%%%%%%%%%%%%%%%%%%%%%%%%%%%%%%%%%%%
Bose-condensates differ from normal gases or fluids  by the existence of
a macroscopic wave-function, their order parameter. The macroscopic
wave-function deeply influences the spectrum of low-lying elementary
excitations in Bose-condensed systems, which become collisionless density
waves with the velocity of sound. In the Bose-Einstein condensates of
alkali-metal vapors in traps these collective sound waves have a discrete
spectrum which is determined by the shapes of the trapping potential and of
the condensate. Many experimental \cite{a1,a2,b} and theoretical
\cite{c1,c2,c3,c4,d,String,Oe,Flie1,Flie2}
studies have been devoted to their study. For Bose-Einstein condensates at
zero temperature which are sufficiently large to validate the Thomas-Fermi
approximation Stringari \cite{String} found an analytical solution for the
sound modes and their eigenfrequencies for the case of spherically symmetric
parabolic traps. In the same work he even determined some of the
eigenfrequencies for axially symmetric anisotropic traps. In subsequent works
\cite{Oe,Flie1,Flie2} the complete solution for the axially symmetric
anisotropic case was given. In particular, it was demonstrated that the
axially symmetric anisotropic problem forms a completely integrable system by
exhibiting explicitly a third conserved operator $\hat{B}$ besides the
wave-operator $\hat{G}$ and the axial angular momentum operator $\hat{L}_z$.
The eikonal or `classical limit' of the sound-waves, determining their
characteristic rays, was also studied in \cite{Flie2} in the axially symmetric
case. In this `classical limit' also the completely anisotropic case of
a triaxial harmonic trap was investigated \cite{Flie2}. It was shown that even
in this case the wave operator in eikonal approximation remains separable in
elliptic coordinates. The complete integrability was demonstrated by exhibiting
three phase-space functions $G$, $B$ and $A$ in involution. However, the
solution for the mode-functions and eigenfrequencies was not yet given in
\cite{Flie2}.

In the present paper we wish to return to this completely anisotropic case.
>From a practical point of view this has become of interest, because the first
completely anisotropic trap has now appeared on the experimental scene
\cite{An}. There the reported trap-frequency ratios are $\omega_1^2:
\omega_2^2:\omega_3^2=1:2:4$. We shall return to this case when we give
a numerical example at the end.

>From a theoretical point of view the problem is also of considerable interest.
The previous results on its classical limit suggest that also the full
wave-operator remains separable in the general anisotropic case. This is indeed
the case, as will be shown here. In fact we shall see that this problem
can be related to a novel class of completely integrable elliptic billiards
on an
inhomogeneously curved space of arbitrary dimensionality. The sound modes
in the Bose-Einstein condensate correspond to the 3-dimensional quantized
version of such a billiard, their characteristics or rays are given by their
classical limit. Furthermore it turns out that the classical limit of the
billiards (in arbitrary dimension) can be connected mathematically to the
equations of motion of an integrable system first studied by C.~Neumann
\cite{Neu} 140 years ago: a mass point on the sphere $|\bbox{x}|=1$ under the
influence of an anisotropic harmonic force.

The problem of collective modes in Bose-Einstein condensates can be viewed as
the problem of small perturbations of the macroscopic wave function around
its static equilibrium. At zero temperature the macroscopic wave
function satisfies the Gross-Pitaevskii equation \cite{GP}
\begin{equation}
\left\{
 -\frac{\hbar^2}{2m}\nabla^2+U(\bbox{x})-\mu+
 \frac{4\pi\hbar^2a_0}{m}|\phi(\bbox{x},t)|^2
\right\}\phi(\bbox{x},t)=i\hbar\dot{\phi}(\bbox{x},t)
\label{eq:1.1}
\end{equation}
where
\begin{equation}
U(\bbox{x})=\frac{1}{2}m\left(\omega_1^2x_1^2+\omega_2^2x_2^2+
\omega_3^2x_3^2\right)
\label{eq:1.2}
\end{equation}
is the anisotropic harmonic potential of the trap, $\mu$ is the chemical
potential, fixed by the requirement that $N=\int d^3x|\phi(\bbox{x},t)|^2$,
and $a_0$ is the $s$-wave scattering length, assumed to be positive throughout
this paper. The macroscopic wave function (at zero temperature) is related to
the number density $n(\bbox{x},t)=|\phi(\bbox{x},t)|^2$ and the
momentum-density $\bbox{g}=\frac{\hbar}{2i}
(\phi^*\bbox{\nabla}\phi-\phi\bbox{\nabla}\phi^*)=
n(\bbox{x})\bbox{v}_s$ which satisfy the euqations of motion
\begin{eqnarray}
\label{eq:1.3}
\frac{\partial n}{\partial t}+\bbox{\nabla}\cdot n \bbox{v_s}&=&0\\
\frac{\partial\bbox{v}_s}{\partial t}+\bbox{\nabla}
\big[
\frac{1}{2}\bbox{v}_s^2 -\frac{\hbar^2}{2m^2}
\frac{\bbox{\nabla}^2\sqrt{n}}{\sqrt{n}}+\frac{1}{m}(U-\mu+
\frac{4\pi\hbar^2a_0}{m}n)\big]&=0&\nonumber
\,.
\end{eqnarray}
We shall assume that the condition
$Na_0\sqrt{m\bar{\omega}/\hbar}\gg1$ is satisfied, where
$\bar{\omega}=(\omega_1\omega_2\omega_3)^{1/3}$, so that the Thomas-Fermi
approximation \cite{TF} can be applied to (\ref{eq:1.3}), where the term
proportional to $(\nabla^2\sqrt{n})/\sqrt{n}$ is neglected. Then the
equilibrium solution with
$\partial n/\partial t=0=\partial\bbox{v}_s/\partial t$ is given by
$\bbox{v}_s=0$, $n=n_0(\bbox{x})=\frac{m}{4\pi\hbar^2a_0}
\big(\mu-U(\bbox{x})\big)$, $\mu=\big(\hbar\bar{\omega}/2\big)
\big(15Na_0/\bar{d}\big)^{2/5}$, where $\bar{d}=\sqrt{\hbar/m\bar{\omega}}$.
Thus the condensate forms a triaxial ellipsoid. The collective excitations in
the same approximation are now determined by eqs.~(\ref{eq:1.3}), linearized
around the equilibrium solution,
\begin{eqnarray}
n
&=&
n_0+\delta n, \quad\bbox{v}_s=\delta\bbox{v}_s\nonumber\\
\label{eq:1.4}
\frac{\partial\delta n}{\partial t}
&+&
\bbox{\nabla}\cdot n_0(\bbox{x})\delta\bbox{v}_s=0\\
\frac{\partial\delta\bbox{v}_s}{\partial t}
&+&
\frac{4\pi\hbar^2a_0}{m^2}\bbox{\nabla}\delta n=0\nonumber\,.
\end{eqnarray}
Eliminating $\delta\bbox{v}_s$ and with the ansatz $\delta n(t)=\psi(\bbox{x})
e^{-i\omega t}$ we are left with the wave equation, first derived along the
present lines by Stringari \cite{String}
\begin{equation}
 \frac{\omega^2}{c_0^2}\psi
 =
 -\bbox{\nabla}\cdot
 (1-\frac{x_1^2}{a^2}-\frac{x_2^2}{b^2}-\frac{x_3^2}{c^2})\bbox{\nabla}\psi\,.
 \label{eq:1.5}
 \end{equation}
Here we introduce the characteristic lengths
\begin{equation}
 a=\sqrt{\frac{2\mu}{m\omega_1^2}}\,,\quad
 b=\sqrt{\frac{2\mu}{m\omega_2^2}}\,,\quad
 c=\sqrt{\frac{2\mu}{m\omega_3^2}}\,,
\label{eq:1.6}
\end{equation}
which are the three semi-axes of the condensate ellipsoid. We note that
\begin{equation}
\omega_1^2 : \omega_2^2 :\omega_3^2 = \frac{1}{a^2} : \frac{1}{b^2} :
\frac{1}{c^2}\,.
\label{eq:1.7}
\end{equation}
We also introduced the velocity of sound $c_0=\sqrt{\mu/m}$ in the center
of the trap. In the following we assume $a^2\ge b^2\ge c^2$ without restriction
of generality, i.e. $\omega_1$ is the smallest of the three trap
frequencies and $\omega_3$ the largest. We shall sometimes use the notation
$a_1=a$, $a_2=b$, $a_3=c$. In this paper we shall be concerned with the
solution of eq.~(\ref{eq:1.5}).

%%%%%%%%%%%%%%%%%%%%%%%%%%%%%%%%%%%%%%%%%%%%%%%%%%%%%
\section{Eikonal approximation}
%%%%%%%%%%%%%%%%%%%%%%%%%%%%%%%%%%%%%%%%%%%%%%%%%%%%%
To get the eikonal approximation to the wave equation (\ref{eq:1.5}) we first
return to its explicitly time-dependent form replacing
$\omega^2\to-\partial^2/\partial t^2$, and replace space and time derivatives
via $i\frac{\partial}{\partial t}\to H$, $-i\hbar\bbox{\nabla}\to\bbox{p}$.
This leaves us with the Hamiltonian
\begin{equation}
H=c_0
\sqrt{p^2
\left(1-\frac{x_1^2}{a^2}-\frac{x_2^2}{b^2}-\frac{x_3^2}{c^2}\right)}
\label{eq:2.1}
\end{equation}
whose trajectories describe the characteristics of the wave equation from which
it was derived. Some features of the classical dynamics described by
eq.~(\ref{eq:2.1}) where studied in ref. \cite{Flie2}. Here we wish to make
a number of additional points.
\begin{enumerate}
\item[a.] {\bf Connection to Neumann's system}\\
     The Hamiltonian equations of motion following from eq.~(\ref{eq:2.1})
     (with time now measured as a length by taking units with $c_0=1$)
     \begin{equation}
     \dot{p}_i
     =
     \frac{x_i}{a_i^2}\,
     \frac{p}{\sqrt{1-\frac{x_1^2}{a^2}-\frac{x_2^2}{b^2}-\frac{x_3^2}{c^2}}}
     \,,\quad
     \dot{x}_i
     =
     \frac{p_i}{p}
     \sqrt{1-\frac{x_1^2}{a^2}-\frac{x_2^2}{b^2}-\frac{x_3^2}{c^2}}
     \label{eq:2.2}
     \end{equation}
     have the interesting property, remarked in \cite{Flie2}, that the dynamics
     of the unit vector $\hat{\bbox{p}}=\bbox{p}/p$ can be decoupled from the
     dynamics of $p$. In fact, eliminating $x_i$ from eqs.~(\ref{eq:2.2}) we
     obtain
     \[
     \dot{\hat{p}}_i=
     \left.\left(
     \frac{x_i}{a_i^2}-\hat{p}_i\sum_j\frac{\hat{p}_jx_j}{a_j^2}
     \right)\right/\sqrt{1-\sum_k x_k^2/a_k^2}
     \]
     and the equations of motion
     \begin{equation}
     \ddot{\hat{p}}_i
     =
     \hat{p}_i
     \left(
     \frac{1}{a_i^2}-\sum_{j=1}^3
     \left(\frac{\hat{p}_j^2}{a_j^2}+\dot{\hat{p}_j^2}\right)
     \right)\,,
     \label{eq:2.3}
     \end{equation}
     with the constraints $\bbox{\hat{p}}^2=1$, $\bbox{\hat{p}}\cdot
     \dot{\hat{\bbox{p}}}=0$. These are formally the equations of motion of
     a particle with unit mass with coordinates $\bbox{\hat{p}}$ on a sphere
     $\bbox{\hat{p}}^2=1$ under the influence of the force $\bbox{F}$ with
     components $F_i=\hat{p}_i/a_i^2$ which were studied by Neumann
     \cite{Neu}.
     A discussion of this problem within the modern mathematical theory of
     integrable systems has been given by Moser \cite{Mo}. He derives its
     conservation laws by constructing a matrix whose eigenvalues are preserved
     under the dynamics (\ref{eq:2.3}). The conserved quantities (see also
     \cite{Cons}) are
     \begin{equation}
     \label{eq:M}
     M_k=-\hat{p}_k^2+\sum_{i\ne k}\frac{a_i^2a_k^2}{a_i^2-a_k^2}
     (\hat{p}_i\dot{\hat{p}}_k-\hat{p}_k\dot{\hat{p}}_i)^2
     \end{equation}
     They satisfy $\sum_k M_k=-\hat{p}^2=1$.
     We note that the
     Hamiltonian $H$ (\ref{eq:2.1}) is no longer among these integrals
     because the absolute value $p$ of the momentum was eliminated in the
     derivation of (\ref{eq:2.3}) from (\ref{eq:2.2}). However the conservation
     of (\ref{eq:2.1}) can be used to recover the motion of $p$ from the
     solution of (\ref{eq:2.3}). There is a new obvious `energy'-integral
     of eq.~(\ref{eq:2.3}) which is given by
     \begin{equation}
     E_N=\frac{1}{2}\sum_i
     \left(\dot{\hat{p}}_i^2-\frac{\hat{p}_i^2}{a_i^2}\right)=
     \frac{1}{2}\sum_i\frac{M_i}{a_i^2}\,.
     \label{eq:2.4a}
     \end{equation}
     It can be expressed in the original $\bbox{x}$, $\bbox{p}$ variables as
     \begin{equation}
     E_N=-\frac{1}{2H^2}\sum_i
     \left\{
     \frac{p_i^2}{a_i^2}
     \left(1-\sum_k\frac{x_k^2}{a_k^2}\right)-
     \frac{x_i^2}{a_i^4}\,p^2+\frac{x_ip_i}{a_i^2}\sum_k
     \frac{x_kp_k}{a_k^2}
     \right\}
     \label{eq:2.4b}
     \end{equation}
     i.e. it is now a quite complicated looking and far from obvious first
     integral of eqs. (\ref{eq:2.2}). With some labor it can be expressed in
     terms of the first integral $A$ which will be introduced in sections 2c
     and 3 (see (\ref{eq:3.15})) as
     \begin{equation}
     E_N=\frac{1}{2}
     \left(\frac{A}{a_1^2a_2^2a_3^2H^2}-\sum_{i=1}^3\frac{1}{a_i^2}\right)
     \label{eq:2.4c}
     \end{equation}
     Another simple linear combination of the $M_k$ is
     \[
     B_N=-\sum_k a_k^2 M_k=
     \sum_ka_k^2\hat{p}_k^2+\frac{1}{2}\sum_{i,k}a_i^2a_k^2
     \big(\hat{p}_i\dot{\hat{p}}_k-\hat{p}_k\dot{\hat{p}}_i\big)^2\,.
     \]
     It can be expressed in terms of the first integral $B$ introduced
     in section 2c and 3 (see (\ref{eq:3.15})) as
     \[
     B_N=\frac{1}{H^2}\big(\sum_k a_k^2p_k^2+
     (\bbox{x}\cdot\bbox{p})\big)^2=\frac{B}{H^2}
     \]
\item[b.] {\bf Connection to a billiard on a curved space}\\
     The Hamiltonian $H$ describes the geodesic motion of a particle in a
     space with the metric
     \begin{equation}
     ds^2=\frac{1}{1-\sum_j\frac{x_j^2}{a_j^2}}\sum_idx_i^2\,.
     \label{eq:2.5}
     \end{equation}
     This metric is inhomogeneous and conformal to the Euclidean metric. Its
     Riemann scalar curvature is given by
     \[
     R=\sum_i\frac{2}{a_i^2}\big(2+\frac{5}{1-\sum_j \frac{x_j^2}{a_j^2}}
     \frac{x_i^2}{a_i^2}\big)
     \]
     and singular on the surface-ellipsoid. The distance to the
     surface-ellipsoid $\sum_j x_j^2/a_j^2=1$ from any point inside the surface
     is finite. Furthermore, the metric velocity $ds/dt$ is constant and
     simply given by $ds/dt=1$ in our present units. The billiard particle
     always reaches the surface-ellipsoid perpendicularly and
     with  diverging orthogonal and finite tangential components
     of the  momentum and is reflected
     with conserved tangential momentum component \cite{Flie2}.
\item[c.] {\bf Separation in ellipsoid coordinates}\\
     Let us introduce elliptical coordinates \cite{WiWa} $\lambda$,$\mu$,
     $\nu$ as the three roots of
     \begin{equation}
     \frac{x_1^2}{a^2+\varrho}+
     \frac{x_2^2}{b^2+\varrho}+
     \frac{x_3^2}{c^2+\varrho}=1
     \label{eq:2.6}
     \end{equation}
     We order these roots $\rho=\lambda ,\mu ,\nu$ according to
     \begin{equation}
     -a^2\le\nu\le -b^2\le\mu\le -c^2\le\lambda\le0\,.
     \label{eq:2.7}
     \end{equation}
     Geometrically surfaces $\lambda=\const$ are ellipsoids, $\mu=\const$ are
     one-sheeted hyperboloids, and $\nu=\const$ are two-sheeted hyperboloids,
     all confocal to the basic ellipsoid
     \[
     \frac{x_1^2}{a^2}+\frac{x_2^2}{b^2}+\frac{x_3^2}{c^2}=1\,.
     \]
     Explicitly, the $x_i$ are given in terms of the new variables by
     \begin{equation}
     x_1=\pm\sqrt{\frac{(a^2+\lambda)(a^2+\mu)(a^2+\nu)}{(a^2-b^2)(a^2-c^2)}}
     \qquad\mbox{\rm and cyclic.}
     \label{eq:2.8}
     \end{equation}
     The Hamiltonian-Jacobi equation
     \begin{equation}
     \omega=H(\bbox{\nabla}S,\bbox{x})
     \label{eq:2.9}
     \end{equation}
     is then separable, as shown in \cite{Flie2}. Written in elliptical
     coordinates it takes the form
     \begin{equation}
     0=\left\{(\mu-\nu)
     \left[
     (a^2+\lambda)(b^2+\lambda)(c^2+\lambda)
      \left(\frac{\partial S}{\partial\lambda}\right)^2
      +\frac{a^2b^2c^2\omega^2}{4\lambda}\right]
      + cyclic \right\}
     \label{eq:2.10}
     \end{equation}
     which is satisfied only if the angular bracket is equal to a linear
     function of $\lambda$, i.e.
     \begin{equation}
     \left[
     (a^2+\lambda)(b^2+\lambda)(c^2+\lambda)
     \left(
     \frac{\partial S}{\partial\lambda}\right)^2
     +\frac{a^2b^2c^2\omega^2}{4\lambda}
     \right]=\frac{1}{4}(-A-B\lambda)\quad\mbox{\rm and cyclic}
     \label{eq:2.11}
     \end{equation}
     where $A$ and $B$ are separation constants.  Eqs.~(\ref{eq:2.11}) can be
     solved
     for $A,B,\omega^2=H^2$ as functions of the coordinates $\lambda$, $\mu$,
     $\nu$ and the canonically conjugate momenta
     $p_\lambda=\partial S/\partial\lambda$, $p_\mu=\partial S/\partial\mu$,
     $p_\nu=\partial S/\partial\nu$. The results, expressed in terms of the
     Cartesian coordinates and momenta have been given in \cite{Flie2} and need
     not to be given here. In any case they are easily recovered from the
     operators $\hat{A}$, $\hat{B}$ derived in section III upon taking the
     classical limit $-i\hbar\bbox{\nabla}\to\bbox{p}$, $\hbar\to0$. Here we
     wish to remark on a direct geometrical significance of the values of the
     three conserved quantities $H^2$, $A$, and $B$. Let us introduce to this
     end the roots $\lambda_1>\lambda_2$
     \begin{equation}
     \lambda_{1,2}=-\frac{A}{2B}\pm\sqrt{\frac{A^2}{4B^2}-\frac{C}{B}}
     \label{eq:2.12}
     \end{equation}
     of the quadratic equation
     \begin{equation}
     B\lambda^2+A\lambda^2+C=0
     \label{eq:2.13}
     \end{equation}
     where $C=a^2b^2c^2\omega^2>0$. The equations for the momenta $p_\lambda$,
     $p_\mu$, $p_\nu$ now take the form
     \begin{equation}
     p_\lambda=\pm
     \sqrt{-\frac{B(\lambda-\lambda_1)(\lambda-\lambda_2)}
     {\lambda(a^2+\lambda)(b^2+\lambda)(c^2+\lambda)}}
     \quad\mbox{\rm and cyclic.}
     \label{eq:2.14}
     \end{equation}
     For a physical motion inside the ellipsoid each of the momenta
     $p_\lambda$, $p_\mu$, $p_\nu$ needs to have an outer and an inner turning
     point. Thus there must be six turning points, which, according to
     eq.~(\ref{eq:2.14}) can only be at values $\lambda_1$, $\lambda_2$,
     $0$, $-a^2$, $-b^2$, $-c^2$. Of these the turning points at $-a^2$,
     $-b^2$, $-c^2$ correspond to coordinate singularities, the turning point
     at 0 is the surface of the condensate and the turning points at
     $\lambda_1$, $\lambda_2$ correspond to caustic surfaces. Therefore
     $\lambda_1$, $\lambda_2$ must be real, and negative in order to be in the
     range of $\lambda$, $\mu$, $\nu$, which imposes the conditions
     \begin{equation}
     A^2>4BC\,,\qquad A>0, \qquad B>0\,.
     \label{eq:2.15}
     \end{equation}
     There are then four possible cases for the roots $\lambda_1$, $\lambda_2$
     in which $p_\lambda$, $p_\mu$, $p_\nu$ are real. These are:
     \begin{enumerate}
     \item[1)] $-a^2<\nu<-b^2<\lambda_2<\mu<-c^2<\lambda_1<\lambda<0$\\
          In this case $p_\lambda$ turns at $p_\lambda=0$ on a surface of inner
	  turning points forming the ellipsoid $\lambda=\lambda_1$ and
	  similarly $p_\mu$ turns at
	  $p_\mu=0$ on a second surface of inner turning points forming a
	  one-sheeted hyperboloid $\mu=\lambda_2$.
     \item[2)] $-a^2<\nu<-b^2<\lambda_2<\mu<\lambda_1<-c^2<\lambda<0$\\
          Here $p_\mu$ turns at an outer surface $\mu=\lambda_1$ and
	  an inner surface $\mu=\lambda_2$ which are both one-sheeted
	  hyperboloids.
     \item[3)] $-a^2<\nu<\lambda_2<-b^2<\mu<-c^2<\lambda_1<\lambda<0$\\
          Here $p_\lambda$ turns on an inner ellipsoid $\lambda=\lambda_1$
	  and $p_\nu$ turns on an outer two-sheeted hyperboloid
	  $\nu=\lambda_2$.
     \item[4)] $-a^2<\nu<\lambda_2<-b^2<\mu<\lambda_1<-c^2<\lambda<0$\\
          Here $p_\mu$ turns on an outer one-sheeted hyperboloid
	  $\mu=\lambda_1$ and $p_\nu$ turns on an outer two-sheeted
	  hyperboloid.
     \end{enumerate}
     The conservation of $A$ and $B$ for given $\omega$ and the particular
     value taken by these quantities thus is reflected in the geometry of
     the two caustic surfaces occuring in each case. Similar results have been
     obtained for billiards in Euclidean space with ellipsoidal boundaries
     \cite{bill}.
\item[d.] {\bf Semiclassical quantum numbers}\\
     A `semiclassical' approach to the soltion of the wave equation could be
     the application of the Bohr-Sommerfeld rule
     \begin{equation}
     I_\lambda=\frac{1}{2\pi}\oint p_\lambda d\lambda=n_\lambda+1/2\quad
     \mbox{\rm and cyclic.}
     \label{eq:2.16}
     \end{equation}
     See ref.~\cite{Cs} for the case of isotropic traps. Here the quantum
     numbers $n_\lambda$, $n_\mu$, $n_\nu$ can be interpreted to count the
     number of
     nodal surfaces with $\lambda=\const$, $\mu=\const$, $\nu=\const$,
     respectively, which are ellipsoids, one-sheeted hyperboloids, and
     two-sheeted hyperboloids
     respectively. The original conserved quantities $A$, $B$ and $\omega^2$
     can be expressed in terms of these quantum numbers and are thereby
     quantized in terms of the three independent integers $n_\lambda$,
     $n_\mu$, $n_\nu$. In section IV these quantum numbers will reappear in
     the exact solution of the wave equation in the slightly different
     notation
     $n_3$, $n_2$, $n_1$, the correspondence being $n_\lambda=n_3$,
     $n_\mu=n_2$, $n_\nu=n_1$.
\end{enumerate}

%%%%%%%%%%%%%%%%%%%%%%%%%%%%%%%%%%%%%%%%%%%%%%%%%%%%%
\section{Separation of the wave equation}
%%%%%%%%%%%%%%%%%%%%%%%%%%%%%%%%%%%%%%%%%%%%%%%%%%%%%
Let us now return to the wave equation
\begin{equation}
\omega^2\psi
 =
 -\bbox{\nabla}\cdot
 (1-\frac{x_1^2}{a^2}-\frac{x_2^2}{b^2}-\frac{x_3^2}{c^2})\bbox{\nabla}\psi\,
\label{eq:3.1}
\end{equation}
where we again adopt units with $c_0=1$. We are interested in the solutions
$\psi$ of eq.~(\ref{eq:3.1}) in the Hilbert space with the scalar product
\begin{equation}
\langle \psi_i|\psi_j\rangle=\int_{TF}d^3x \psi^*_i(\bbox{x})\psi_j(\bbox{x})\,
\label{eq:3.1a}
\end{equation}
where the integration is extended over the interior of the Thomas-Fermi
ellipsoid. With this choice of the scalar product the operator $\hat{G}=
-\bbox{\nabla}\cdot(1-\sum_ix_i^2/a_i^2)
\bbox{\nabla}$ is self-adjoint if we pose as boundary
condition
\[
 \left(1-\sum_i\frac{x_i^2}{a_i^2}\right)\partial_n\psi=0
\]
on the boundary, i.e. the normal derivative $\partial_n\psi$ should diverge
there less than
$(1-\sum_i x_i^2/a_i^2)^{-1}$. This can be satisfied by choosing the Hilbert
space of polynomials of finite order with the scalar product (\ref{eq:3.1a}).
These polynomials can be divided into eight different parity classes depending
on whether they are even or odd under the inversion of any of the three
coordinates $x_1$, $x_2$, $x_3$. To be specific we put
\[
  \psi=x_1^\alpha x_2^\beta x_3^\gamma\,P_m(x_1^2,x_2^2,x_3^2)
\]
where $\alpha,\beta,\gamma=0,1$ determine the parity class and $P_m$ is a
polynomial of order $m$ in $x_1^2$, $x_2^2$, $x_3^2$. Eq.~(\ref{eq:3.1})
is now transformed to elliptical coordinates where it takes the form
\begin{eqnarray}
a^2b^2c^2\omega^2=&-\frac{4\lambda\mu\nu}{(\lambda-\mu)(\mu-\nu)(\nu-\lambda)}
\left\{(\mu-\nu)\big [F(\lambda)\frac{\partial^2}{\partial\lambda^2}+
(\frac{F(\lambda)}{\lambda}+\frac{1}{2}F'(\lambda))
\frac{\partial}{\partial\lambda}\big]+ cyclic \right\}\psi \,
\label{eq:3.2}
\end{eqnarray}
with
\begin{equation}
 F(\varrho)=(a^2+\varrho)(b^2+\varrho)(c^2+\varrho)\,.
\label{eq:3.3}
\end{equation}
Eq.~(\ref{eq:3.2}) can be solved by separation of variables via
\begin{equation}
 \psi=\varphi_\lambda(\lambda)\varphi_\mu(\mu)\varphi_\nu(\nu)
\label{eq:3.4}
\end{equation}
which leads to
\begin{equation}
 0=(\mu-\nu)g_\lambda(\lambda)+(\nu-\lambda)g_\mu(\mu)+(\lambda-\mu)
  g_\nu(\nu)
\label{eq:3.5}
\end{equation}
with
\begin{equation}
g_{\varrho}(\varrho)=\frac{1}{\varphi_\varrho(\varrho)}
\big[-F(\varrho)\frac{d^2}{d\varrho^2}+
\big(\frac{F(\varrho)}{\varrho}+\frac{1}{2}F'(\varrho)\big)
\frac{d}{d\varrho}+\frac{a^2b^2c^2\omega^2}{4\varrho}\big]
\varphi_\varrho(\varrho)\,
\label{eq:3.6}
\end{equation}
If eq.~(\ref{eq:3.5}) is rewritten as
\begin{equation}
 g_\lambda (\lambda)=
  \frac{\mu g_\nu(\nu)-\nu g_\mu(\mu)}{\mu-\nu}+
   \lambda\frac{g_\mu(\mu)-g_\nu(\nu)}{\mu-\nu}
\label{eq:3.7}
\end{equation}
it becomes apparent that it can hold as a identity in $\lambda$, $\mu$, $\nu$
only if
\begin{eqnarray}
g_\lambda(\lambda)&=&-\frac{A}{4}-\frac{B}{4}\lambda\nonumber\\
\label{eq:3.8}
g_\mu(\mu)&=&-\frac{A}{4}-\frac{B}{4}\mu\\
g_\nu(\nu)&=&-\frac{A}{4}-\frac{B}{4}\nu\nonumber\,
\end{eqnarray}
where $A$ and $B$ are the same constants in all three of eqs.~(\ref{eq:3.8}).
>From eq.~(\ref{eq:3.8}) and (\ref{eq:3.6}) the three constants $\omega^2$,
$A$, $B$ can be expressed as eigenvalues of certain corresponding operators
$\hat{G}$, $\hat{A}$, $\hat{B}$. To do this explicitly we define the operator
$\hat{F}_\varrho$ for arbitrary $\varrho$ as
\begin{equation}
\label{eq:3.9}
\hat{F}_\varrho\psi=4\big[\varrho F(\varrho)
\frac{\partial^2}{\partial\varrho^2}+(F(\varrho)+
\frac{1}{2}\varrho F'(\varrho))
\frac{\partial}{\partial\varrho}\big]\psi\,
\end{equation}
because then eqs.~(\ref{eq:3.8})  with (\ref{eq:3.6}) can be rewritten simply
as
\begin{equation}
 \hat{F}_\varrho\psi=(a^2b^2c^2\omega^2+A\varrho+B\varrho^2)\psi
\label{eq:3.10}
\end{equation}
where $\psi$ is the total wave function and $\varrho=\lambda,\mu,\nu$.
Solving for $\omega^2\psi$, $A\psi$, $B\psi$ we
find the simultaneous eigenvalue equations for $\omega^2$, $B$, $A$, namely
eq.~(\ref{eq:3.2}) which we abbreviate as $\omega^2\psi=\hat{G}\psi$ and
\begin{equation}
 \hat{B}\psi=B\psi\,,\qquad \hat{A}\psi=A\psi
\label{eq:3.11}
\end{equation}
with
\begin{eqnarray}
\label{eq:3.12}
\hat{B}&=&-\frac{4}{(\lambda-\mu)(\mu-\nu)(\nu-\lambda)}
\left\{(\mu-\nu)\big[\lambda F(\lambda)\frac{\partial^2}{\partial\lambda^2}+
(F(\lambda)+\frac{1}{2}\lambda F'(\lambda))\frac{\partial}{\partial\lambda}
\big] + cyclic \right\}\\
\hat{A}&=&\frac{4}{(\lambda-\mu)(\mu-\nu)(\nu-\lambda)}\left\{(\mu^2-\nu^2)
\big[\lambda F(\lambda)\frac{\partial^2}{\partial\lambda^2}+(F(\lambda)+
\frac{1}{2}\lambda F'(\lambda))\frac{\partial}{\partial\lambda}\big] +
cyclic \right\}
\end{eqnarray}
A lengthy but straight-forward calculation then yields the operators $\hat{A}$
and $\hat{B}$ in Cartesian coordinates
\begin{eqnarray}
\label{eq:3.14}
\hat{A}&=&\left\{\big[(b^2+c^2)(x_1^2-a^2)+a^2(x_2^2+x_3^2)\big]
\frac{\partial^2}{\partial x_1^2}+2a^2x_2x_3
\frac{\partial^2}{\partial x_2\partial x_3}+3(b^2+c^2)x_1
\frac{\partial}{\partial x_1}
+ cyclic \right\}\\
\label{eq:3.15}
\hat{B}&=&(\bbox{x\cdot\nabla})(\bbox{x\cdot\nabla}+3)-a^2\frac{\partial^2}
{\partial x_1^2}-b^2\frac{\partial^2}{\partial x_2^2}-
c^2\frac{\partial^2}{\partial x_3^2}
\end{eqnarray}
By construction the eigenvalue equations for $\hat{A}$, $\hat{B}$ and $\hat{G}$
can be satisfied simultaneously, i.e. these operators must commute, as one
may also check by explicit calculation
\[
 [\hat{G},\hat{A}]=[\hat{B},\hat{A}]=[\hat{G},\hat{B}]=0\,.
\]
In the axially symmetric case, e.g. $b^2=c^2$, the operator $\hat{A}$ may be
expressed in terms of the angular momentum $\hat{L}_z$ around the axis of
symmetry, here chosen as the 1-axis, according to
\[
 \hat{A}=c^2\hat{B}+(a^2-c^2)\hat{L}_z^2+a^2c^2\hat{G}\,.
\]

%%%%%%%%%%%%%%%%%%%%%%%%%%%%%%%%%%%%%%%%%%%%%%%%%%%%%
\section{Solution of the wave equation}
%%%%%%%%%%%%%%%%%%%%%%%%%%%%%%%%%%%%%%%%%%%%%%%%%%%%%
After the separation of variables the equation to be solved follows from
eq.~(\ref{eq:3.6}) with (\ref{eq:3.8}) as
\begin{equation}
\big[-\varrho F(\varrho)\frac{d^2}{d\varrho^2}-(F(\varrho)+
\frac{1}{2}\varrho F'(\varrho))\frac{d}{d\varrho}+
\frac{1}{4}(a^2b^2c^2\omega^2+A\varrho +B\varrho^2)\big]
\varphi_\varrho(\varrho)=0
\label{eq:4.1}
\end{equation}
for $\varrho=\lambda,\mu,\nu$, i.e. precisely the same equation appears in all
three elliptical coordinates. We now restrict the solutions of (\ref{eq:4.1})
to the space of polynomials in Cartesian coordinates. This means that in
elliptical coordinates they must be of the form
\begin{equation}
 \varphi_\varrho(\varrho)
  =
  |a^2+\varrho|^{\frac{\alpha}{2}}
  |b^2+\varrho|^{\frac{\beta}{2}}
  |c^2+\varrho|^{\frac{\gamma}{2}}P_m(\rho)
\label{eq:4.2}
\end{equation}
where $P_m(\varrho)$ is a polynomial of order $m$ and the exponents $\alpha$,
$\beta$, $\gamma$ can take on the values 0 and 1. Inserting this ansatz in
eq.~(\ref{eq:4.1}) results in an expression containing the same
prefactors as eq.~(\ref{eq:4.2}) but multiplied with a polynomial of order
$m+2$ whose coefficients must all vanish, yielding $m+2$ equations from which
the three eigenvalues $A, B, \omega^2$ and the $m-1$ unknown coefficients of
$P_m$ (one coefficient is fixed by normalization) must be determined. The
vanishing of the coefficient of highest order fixes the value of $B$ as
\begin{equation}
B=(2m+\alpha+\beta+\gamma)(2m+\alpha+\beta+\gamma+3)\,.
\label{eq:4.3}
\end{equation}
The condition that the coefficient of the next to highest order term vanishes
in principle fixes the value of $A$. However, it turns out that $A$, and also
$\omega^2$, cannot be determined without at the same time calculating all the
coefficients of the polynomial $P_m$ in eq.~(\ref{eq:4.2}). Simple results
are therefore not obtained in this way, except for a few of the lowest lying
modes.

Therefore a different procedure is used. It is a generalization of the analysis
used in the solution of the Lam\'e equation \cite{WiWa}. We shall assume now,
and make plausible at the end of this section that the eigenvalues of the three
separation constants $\omega^2$, $A$, $B$ uniquely specify the corresponding
eigenfunction, up to an arbitrary multiplicative factor. Then one can choose
\begin{equation}
 \varphi_\lambda (\varrho)=\varphi_\mu(\varrho)=\varphi_\nu(\varrho)\,.
\label{eq:4.4}
\end{equation}
In the polynomial ansatz (\ref{eq:4.2}) we write
\begin{equation}
P_m(\varrho)=\prod_{i=1}^m(\varrho-\theta_i)
\label{eq:4.5}
\end{equation}
where $\theta_i$ are the (possibly complex) roots of the polynomial $P_m$. The
ansatz for the total wave function
$\psi=\varphi_\lambda(\lambda)\varphi_\mu(\mu)\varphi_\nu(\nu)$ then becomes
\begin{eqnarray}
\psi= const &|&(a^2+\lambda)(a^2+\mu)(a^2+\nu)|^{\alpha/2}|
(b^2+\lambda)(b^2+\mu)(b^2+\nu)|^{\beta/2}\nonumber\\
\dot &|&(c^2+\lambda)(c^2+\mu)(c^2+\nu)|^{\gamma/2}
\prod_{i=1}^{m}(\lambda-\theta_i)(\mu-\theta_i)(\nu-\theta_i)
\label{eq:4.6}
\end{eqnarray}
Using eqs.~(\ref{eq:2.8}) and the identity
\begin{equation}
 \frac{x_1^2}{a^2+\theta_i}+\frac{x_2^2}{b^2+\theta_i}+
  \frac{x_3^2}{c^2+\theta_i}-1
  =
   \frac{(\lambda-\theta_i)(\mu-\theta_i)(\nu-\theta_i)}
    {(a^2+\theta_i)(b^2+\theta_i)(c^2+\theta_i)}
\label{eq:4.7}
\end{equation}
the wave function (\ref{eq:4.6}) can be written rather simply in Cartesian
coordinates as
\begin{equation}
 \psi(\bbox{x})= x_1^\alpha x_2^\beta x_3^\gamma\prod_{i=0}^m
  \left(\frac{x_1^2}{a^2+\theta_i}+\frac{x_2^2}{b^2+\theta_i}+
   \frac{x_3^2}{c^2+\theta_i}-1\right)
\label{eq:4.8}
\end{equation}
where for $i=0$ the factor under the product is defined as 1. In  
(\ref{eq:4.8}) and in the following we omitt a normalization constant and work  
with unnormalized wave functions. They
are completely parametrized by the yet unknown parameters $\theta_i$
which determine the nodal surfaces. As can be seen directly from  
(\ref{eq:4.7}) the remarkably simple form of the wave function (\ref{eq:4.8})  
is a direct consequence of the separation ansatz. For all of the following  
considerations the form (\ref{eq:4.8}) of the wave function will be used. It  
follows from (\ref{eq:4.8}) that the
nodal surfaces are quadrics confocal to the Thomas-Fermi ellipsoid. In order
to relate the eigenvalues of $\hat{G}$, $\hat{A}$, $\hat{B}$ to the $\theta_i$
we insert the ansatz (\ref{eq:4.8}) in the eigenvalue equations. The
calculations become simpler with the use of the intermediate variables
\begin{eqnarray}
\varphi_i &=&
 \frac{x_1^2}{a^2+\theta_i}+\frac{x_2^2}{b^2+\theta_i}+
  \frac{x_3^2}{c^2+\theta_i}-1\nonumber\\
 \label{eq:4.9}
\varphi_0 &=&
  \frac{x_1^2}{a^2}+\frac{x_2^2}{b^2}+\frac{x_3^2}{c^2}-1\\
\Pi &=& \prod_i^m\varphi_i\nonumber
\end{eqnarray}
Then
\begin{equation}
 \frac{\partial\Pi}{\partial\varphi}\varphi_i
  =
  \Pi\,,\qquad \frac{\partial^2\Pi}{\partial\varphi_i\partial\varphi_j}
   \varphi_j=\frac{\partial\Pi}{\partial\varphi_i}\,.
\label{eq:4.10}
\end{equation}
>From $\hat{G}\psi=\omega^2\psi$ with $\psi=x_1^\alpha x_2^\beta x_3^\gamma
\Pi$ we obtain
\begin{eqnarray}
\label{eq:4.11}
\hat{G}\psi=x_1^\alpha x_2^\beta x_3^\gamma \Bigg\{
&& [\frac{2\alpha}{a^2}+\frac{2\beta}{b^2}+\frac{2\gamma}{c^2}-4
\sum_{i=1}^m\frac{1}{\theta_i} ] \Pi\nonumber\\
&& +\varphi_0\sum_{i=1}^m\frac{\partial \Pi}{\partial\varphi_i}
\big[\frac{4}{\theta_i}+\frac{4\alpha +2}{a^2+\theta_i}+
\frac{4\beta +2}{b^2+\theta_i}+\frac{4\gamma +2}{c^2+\theta_i}+
\mathop{\mathord{\sum}'}_{j=1}^m\frac{8}{\theta_i-\theta_j}\big]\Bigg\}
   \nonumber \,.
\end{eqnarray}
Thus the $\theta_i$ have to satisfy
\begin{equation}
\label{eq:4.12}
G_i(\theta)=\frac{4}{\theta_i}+\frac{4\alpha+2}{a^2+\theta_i}+
\frac{4\beta+2}{b^2+\theta_i}+\frac{4\gamma+2}{c^2+\theta_i}+
\mathop{\mathord{\sum}'}_j\frac{8}{\theta_i-\theta_j}=0
\end{equation}
for $i=1,2,\dots,m$. Here and in the following $\mathop{\mathord{\sum}'}$
denotes the sum without
the diagonal term. The eigenvalues of $\hat{G}$ become
\begin{equation}
 \frac{\omega^2}{c_0^2}=\frac{2\alpha}{a^2}+\frac{2\beta}{b^2}+
 \frac{2\gamma}{c^2}-4
  \sum_{i=1}^m\frac{1}{\theta_i}\,.
\label{eq:4.13}
\end{equation}
(where we momentarily restored $c_0^2$ for later convenience). For $m=0$ the
sum $\sum_{i=1}^m1/\theta_i$ has to be interpreted as 0.
Similarly we obtain for $\hat{B}$
\begin{eqnarray}
\hat{B}\psi=x_1^\alpha x_2^\beta x_3^\gamma
\Bigg\{(2m+\alpha+\beta+\gamma)(2m+\alpha+\beta+\gamma+3)\Pi+
\sum_{i=1}^m\frac{\partial\Pi}{\partial\varphi_i}\theta_i G_i(\theta)\Bigg\}
\label{eq:4.14}
\end{eqnarray}
which leads again to the condition $G_i(\theta)=0$ and otherwise gives back
the
eigenvalue (\ref{eq:4.3}) for $B$, which, in particular turns out to be
independent of the $\theta_i$. Finally applying $\hat{A}$ to $\psi$ we obtain
\begin{eqnarray}
\label{eq:4.15}
\hat{A}\psi=x_1^\alpha x_2^\beta x_3^\gamma \Bigg\{
&& [2\beta\gamma a^2+2\gamma\alpha b^2+2\alpha\beta c^2\nonumber\\
&& +(4m+3)\big(\alpha(b^2+c^2)+\beta(c^2+a^2)+\gamma(a^2+b^2)\big)\nonumber\\
&& +4(a^2+b^2+c^2)m(m+1)\\
&& +(4\alpha+4\beta+4\gamma+8m+2)\sum_{i=1}^m\theta_i ] \Pi\nonumber\\
&& +\sum_{i=1}^m\frac{\partial\Pi}{\partial\varphi_i}
   \left[(a^2+b^2+c^2-x_1^2-x_2^2-x_3^2)\theta_i+\theta_i^2\right]G_i(\theta)
   \Bigg\}
   \nonumber \,.
\end{eqnarray}
Again $G_i(\theta)=0$ must be satisfied, and the eigenvalues of $\hat{A}$ then
can be read off eq.~(\ref{eq:4.15}) in the form
\begin{eqnarray}
\label{eq:4.16}
A= &4& (\alpha+\beta+\gamma+2m+\frac{1}{2})\sum_i^m\theta_i\nonumber\\
&+& [2\alpha\beta c^2+(4m+3)\alpha(b^2+c^2)+4m(m+1)a^2+ cyclic].
\end{eqnarray}
It remains to determine the $\theta_i$ from the $m$ equations (\ref{eq:4.12})
$G_i(\theta)=0$. First we show that all $\theta_i$ are real. To prove this
let us suppose they are complex, in which case they also satisfy
$G_i(\theta^*)=0$.
Hence
\begin{equation}
 0=\sum_i(\theta_i-\theta_i^*) (G_i(\theta)-G_i(\theta^*))
\label{eq:4.17}
\end{equation}
which, after some algebra, and writing $\theta_i=|\theta_i|e^{i\vartheta_i}$,
leads to
\begin{eqnarray}
0 &=&\sum_i(\frac{4}{|\theta_i|^2}+\frac{4\alpha+2}{|a^2+\theta_i|^2}+
\frac{4\beta+2}{|b^2+\theta_i|^2}+
\frac{4\gamma+2}{|c^2+\theta_i|^2})|\theta_i|^2\sin^2\vartheta_i\nonumber\\
&+& \mathop{\mathord{\sum}'}_{ij}\frac{4}{|\theta_i-\theta_j|^2}
(|\theta_i|\sin\vartheta_i-|\theta_j|\sin\vartheta_j)^2
\label{eq:4.18}
\end{eqnarray}
This condition is only satisfied if all phases  $\vartheta_i$ equal either
0 or $\pi$, i.e. if the $\theta_i$ are real. A permutation of two $\theta_i$
does not lead to a new eigenfunction. Hence the $\theta_i$ can be assumed
ordered according to
\begin{equation}
 \theta_m\le\theta_{m-1}\le\dots\le\theta_1\,.
\label{eq:4.19}
\end{equation}
Let us consider the cases $m=0$, $m=1$. For $m=0$ we obtain from
(\ref{eq:4.13}) eight
eigenfrequencies
\begin{equation}
 \omega^2=c_0^2\left(
 \frac{2\alpha}{a^2}+\frac{2\beta}{b^2}+\frac{2\gamma}{c^2}
 \right)
\label{eq:4.20}
\end{equation}
for $\alpha,\beta,\gamma=0,1$. These are the frequencies
$\omega^2=0,\omega_1^2,
\omega_2^2,\omega_3^2,\omega_1^2+\omega_2^2,\omega_1^2+\omega_3^2,
\omega_2^2+\omega_3^2, \omega_1^2+\omega_2^2+\omega_3^2$. For $m=1$ one has to
solve just a single cubic equation
\begin{equation}
0=\frac{4}{\theta_1}+\frac{4\alpha+2}{a^2+\theta_1}+
 \frac{4\beta+2}{b^2+\theta_1}+
  \frac{4\gamma+2}{c^2+\theta_1}
\label{eq:4.21}
\end{equation}
to find three different  values for $\theta_1$, which we order according to
\begin{equation}
 -a^2<\theta_{1,3}<-b^2<\theta_{1,2}<-c^2<\theta_{1,1}<0
\label{eq:4.22}
\end{equation}
which correspond, for each of the eight choices of the tripel
$(\alpha,\beta,\gamma)$, to three independent solutions. Taken together the
case $m=1$ therefore gives 24 different frequencies
\begin{equation}
 \omega^2=\alpha\omega_1^2+\beta\omega_2^2+\gamma\omega_3^2-
  \frac{4c_0^2}{\theta_{1,k}}\qquad k=1,2,3\,.
\label{eq:4.23}
\end{equation}
For the special case $\alpha,\beta,\gamma$ all vanishing the result
(\ref{eq:4.23}) with the cubic equation (\ref{eq:4.21}) was already obtained
by Stringari \cite{rev}. The three roots $\theta_{1,k}$ correspond to wave
functions with a single nodal
surface, which for $k=3$ is a two-sheeted hyperboloid, for $k=2$ a one-sheeted
hyperboloid and for $k=1$ is an ellipsoid.

Let us now turn to the case of general order $m$ of the polynomial mode
function. It is quite remarkable that in this case the $m$ equations
$G_i(\theta)=0$ can be derived as the extrema of a single potential
$V(\theta)$
\begin{equation}
G_i(\theta)=-\frac{\partial V(\theta)}{\partial\theta_i}=0
\label{eq:4.24}
\end{equation}
where
\begin{eqnarray}
8V(\theta)= -\frac{1}{2}\sum_{i=1}^m\ln|\theta_i| &-& (\frac{\alpha}{2}+
\frac{1}{4})\sum_i^m\ln|a^2+\theta_i|-(\frac{\beta}{2}+
\frac{1}{4})\sum_i^m\ln|b^2+\theta_i|\nonumber\\
&-& (\frac{\gamma}{2}+\frac{1}{4})\sum_i^m\ln|c^2+\theta_i|
- \sum_{i=1}^m\sum_{j=i+1}^{m}\ln|\theta_i-\theta_j|
\label{eq:4.25}
\end{eqnarray}
Thus the problem has now become the following exercise in statics: In a
one-dimensional space one
has four fixed positive fictitious point-charges aligned along the negative
$\theta$-axis, namely

\begin{tabular}{c@{\quad}l@{\quad}c@{\quad}l}
 $a$ charge & $1/2$ & at & $\theta=0$,\\
 $a$ charge & $\frac{\alpha}{2}+1/4$ & at & $\theta=-a^2$,\\
 $a$ charge & $\frac{\beta}{2}+1/4$ & at & $\theta=-b^2$,\\
 $a$ charge & $\frac{\gamma}{2}+1/4$ & at & $\theta=-c^2$,
\end{tabular}
\vskip0.25cm
\noindent
between these fixed charges $m$ movable positive point-charges of unit
strength, and interacting among themselves and with the fixed charges with
1-dimensional inverse-distance forces, are to
be distributed in such a way that a force-equilibrium (\ref{eq:4.24}) may
result. It is clear from the form of the potential (\ref{eq:4.25})
that the movable charges can be distributed arbitrarily over the three
intervals, namely

\begin{tabular}{cc@{\quad}l}
 $n_3$ charges & in & $-c^2<\theta<0$,\\
 $n_2$ charges & in & $-b^2<\theta<-c^2$,\\
 $n_1$ charges & in & $-a^2<\theta<-b^2$
\end{tabular}
\vskip0.25cm
\noindent
with $n_1+n_2+n_3=m$. There are ${m+2\choose 2}=\frac{(m+2)!}{m!2!}$ ways to
make this distribution, each leading to a different unique equilibrium
configuration for the movable charges at positions $\theta_i$. It is
immediately clear from the mechanical analogy that all $\theta_i$ must be
different. The three
integer numbers $n_1$, $n_2$, $n_3$ are therefore the natural quantum numbers
of the problem. As the eigenvalue $B$ is independent of the positions of the
movable charges, it must be ${m+2\choose 2}$-fold degenerate for each of the
eight choices of the triple $(\alpha,\beta,\gamma)$. The uniqueness of the
equilibrium distribution  of the $\theta_i$ for given $n_1$, $n_2$, $n_3$,
which determine uniquely the eigenvalues $\omega^2$, $A$, $B$, justifies
a posteriori the uniqueness assumption made after eq.~(\ref{eq:4.3}). It is
very easy to find the minima $\theta_i$ of the potential (\ref{eq:4.25})
numerically for any desired triple of integers $n_1$, $n_2$, $n_3$ and to
determine thereby the mode frequencies (\ref{eq:4.13}) and the corresponding
mode functions (\ref{eq:4.8}). In table 1 we give the 20 lowest
lying mode frequencies for the experimentally realized case $\omega_1^2:
\omega_2^2:\omega_3^2=1:2:4$ in units of the smallest trap frequency
$\omega_1$. Also  given there are the parities $\alpha$, $\beta$, $\gamma$
(e.g. there
is even parity in $x_1$ if $\alpha=0$ and odd if $\alpha=1$), and the quantum
numbers $n_3$, $n_2$, $n_1$. The latter give, respectively, the numbers of
nodal surfaces of ellipsoidal, one-sheeted hyperbolic and two-sheeted
hyperbolic form, all
confocal to the Thomas-Fermi ellipsoid. The ellipsoidal nodal surfaces counted
by $n_3$ are more elongated ellipsoids inside the Thomas-Fermi ellipsoid and
would correspond to radial waves in the spherically symmetric case. As the
$\theta$-values counted by $n_3$ are in the group $-c^2<\theta<0$ with the
smallest absolute values, it is clear from (\ref{eq:4.13}) that the ellipsoidal
nodal surfaces lead to the highest frequencies. The one-sheeted hyperbolic
nodal surfaces counted by $n_2$ are ellipsoidal hyperboloids around the
$x_3$-axis which intersect planes $x_3 =\const$ in ellipses and planes
$x_2 =\const\cdot x_1$ in the two branches of hyperbolas, the hyperbolas
opening up
in all directions orthogonal to the $x_3$-axis. Finally, the two-sheeted
hyperbolic nodal surfaces counted by $n_1$ are formed by the two branches of
ellipsoidal hyperboloids around the $1$-axis opening up in the positive and
negative $x_1$-direction. They are cut in ellipses by planes $x_1=\const$ and
in
hyperbolas by planes $x_2=\const$ and $x_3=\const$. The quantum numbers $n_2$
and $n_1$ are clearly the analogues of angular momentum quantum numbers (i.e.
the quantum numbers of spherical harmonics) for the elliptic geometry.

%%%%%%%%%%%%%%%%%%%%%%%%%%%%%%%%%%%%%%%%%%%%%%%%%%%%%
\section{Symmetric traps as limiting cases}
%%%%%%%%%%%%%%%%%%%%%%%%%%%%%%%%%%%%%%%%%%%%%%%%%%%%%
The cases of axially symmetric and isotropic traps must of course be
contained in our results as limiting cases. Let us see how.
\begin{enumerate}
\item[a.] {\bf Axially-symmetric trap}\\
     Let us suppose we have axial symmetry of the trap around the $x_1$-axis.
     In this case we should study the limit $b\to c$ from above. This limit
     has to be taken in the expression for the wave function (\ref{eq:4.8}),
     in the force equation (\ref{eq:4.12}) and in the result for the mode
     frequencies (\ref{eq:4.13}). The positions $\theta_i$ of the $n_2$
     charges
     which have been distributed in the interval $-b^2<\theta<-c^2$ all
     approach $-c^2$ in
     the limit. Therefore the mode frequencies in the limit become
     \begin{equation}
     \frac{\omega^2}{c_0^2}
     =
     \frac{2\alpha}{a^2}+\frac{2\beta+2\gamma+4n_2}{c^2}-
     4\left(\sum_{i=1}^{n_1}+\sum_{i=n_1+n_2+1}^m\right)\frac{1}{\theta_i}\,.
     \label{eq:5.1}
     \end{equation}
     The force-equilibrium for the $n_1+n_2$ charges outside the interval
     $[-b^2,-c^2]$, which alone enter the sums in eq.~(\ref{eq:5.1}), becomes
     \begin{equation}
     0=\frac{4}{\theta_i}+\frac{4\alpha+2}{a^2+\theta_i}+
     \frac{4\beta+4\gamma+8n_2+4}{c^2+\theta_i}+
	    \left(\mathop{\mathord{\sum}'}_{j=1}^{n_1}+
	    \mathop{\mathord{\sum}'}_{j=n_1+n_2+1}^{m}\right)
     \frac{8}{\theta_i-\theta_j}
     \label{eq:5.2}
     \end{equation}
     for $i=1,2,\dots,n_1$; $n_1+n_2+1,\dots,m$. The solution of
     eq.~(\ref{eq:5.2}) is sufficient to determine the mode frequencies.
     However, the wave functions still depend on the asymptotics of the
     distribution of the $n_2$ charges in $-b^2<\theta<-c^2$. The force
     equations for $i=n_1+1$, $n_1+2,\dots,n_1+n_2$ have of course to be
     handled with some care, as the interaction terms between these charges,
     which approach each other arbitrarily closely, diverge. To isolate and
     divide out the diverging prefactor, which would be automatically
     cancelled
     if we worked with normalized wave functions, we define parameters $t_i$
     with $-1\le t_i\le0$ by $\theta_{n_1+i}=-c^2+t_i(b^2-c^2)$. Then the
     limiting term of the force equations in question for $b\to c$ reads
     \begin{equation}
     0=\frac{4\gamma +2}{t_i}+\frac{4\beta+2}{1+t_i}+
     {\mathop{\mathord{\sum}'}^{n_2}_{j=1}}\frac{8}{t_i-t_j}\quad i=1,2,\dots,n_2\,.
     \label{eq:5.3}
     \end{equation}
     These equations can now be solved by similar techniques as used before.
     Fortunately, however, it will be sufficient to use (\ref{eq:5.3}) without
     explicit knowledge of its solutions. Turning now to the wave functions
     we define cylinder coordinates around the $x_1$-axis by $x_1=z$,
     $x_2=\varrho\sin\varphi$, $x_3=\varrho\cos\varphi$. Using the variables
     $t_i$ and multiplying by a factor $(b^2-c^2)^{n_2}$ to eliminate
     the divergence the dominant term of the wave function for $b\to c$ becomes
     \begin{eqnarray}
      \psi
      &\sim &
      x_1^\alpha x_2^\beta x_3^\gamma\prod_{i=1}^m
     \left(\frac{x_1^2}{a^2+\theta_i}+
     \frac{x_2^2}{b^2+\theta_i}+
     \frac{x_3^2}{c^2+\theta_i}\right)\nonumber\\
     \label{eq:5.4}
     &&\\
     &\sim &
     z^\alpha\varrho^{2n_2+\beta+\gamma}\sin^\beta\varphi\cos^\gamma\varphi
     \prod_{j=1}^{n_2}(\cos^2\varphi+t_j)
     \prod_{i=1}^{n_1}\prod_{i=m-n_3+1}^m
     \left(
     \frac{\varrho^2}{c^2+\theta_i}+\frac{z^2}{a^2+\theta_i}-1
     \right)\nonumber\,.
     \end{eqnarray}
     Using the force equilibrium (\ref{eq:5.3}) for the $n_2$ charges pinned
     between $-b^2$ and $-c^2$ it can be shown that
     \begin{equation}
     \sin^\beta\varphi\cos^\gamma\varphi \prod_{i=1}^{n_2}(\cos^2\varphi+t_i)
     \sim
     \cos\left[(2n_2+\beta+\gamma)\varphi-\frac{\beta\pi}{2}\right]\,.
     \label{eq:5.5}
     \end{equation}
     As a result the wave function (\ref{eq:5.4}) for $b\to c$ goes over to
     \begin{eqnarray}
     \psi
     &\sim&
     \varrho^{2n_2+\beta+\gamma}\cos
     \left[(2n_2+\beta+\gamma)\varphi-\frac{\pi\beta}{2}\right]\nonumber\\
     \label{eq:5.6}
     &&\\
     && \quad \cdot
     z^\alpha\prod_{i=1}^{n_1}\prod_{i=m-n_3+1}^m
     \left(
     \frac{\varrho^2}{c^2+\theta_i}+\frac{z^2}{a^2+\theta_i}+1
     \right)\nonumber\,.
     \end{eqnarray}
     The quantum number $\ell_z$ for the conserved angular momentum around the
     $z$-axis can be read off from eq.~(\ref{eq:5.6})
     \[
     |\ell_z|=2n_2+\beta+\gamma\,.
     \]
     Clearly one may take linear combinations of the wave functions
     (\ref{eq:5.6}) for $|\ell_z|$ fixed but suitably changing $\beta$,
     $\gamma$ or $n_2$ to form eigenstates $\sim e^{\pm i|\ell_z|\varphi}$
     of $L_z=-i \partial/\partial\varphi$.
\item[b.] {\bf Isotropic traps:}\\
     We can finally take the further limit $c\to a$ in the results of the
     preceding section. By similar considerations as described there we obtain
     for the mode frequencies
     \begin{equation}
     \frac{\omega^2}{c_0^2}
     =
     \frac{2\alpha+2\beta+2\gamma+4n_1+4n_2}{a^2}-
     4\sum_{i=m-n_3+1}^m\frac{1}{\theta_i}\,.
     \label{eq:5.7}
     \end{equation}
     The positions of the $n_3$ free charges satisfy the force equilibrium
     \begin{eqnarray}
     0 &=& \frac{4}{\theta_i}+
     \frac{4\alpha+4\beta+4\gamma+8n_1+8n_2+6}{a^2}+'
     \nonumber\\
     \label{eq:5.8}
     && \\
     &&\quad +\mathop{\mathord{\sum}'}^m_{j=m-n_3+1}
     \frac{8}{\theta_i-\theta_j}
     \quad\mbox{\rm for}\ i=m-n_3+1,\dots,m\nonumber\,.
     \end{eqnarray}
     Introducing $\varrho=r\sin\vartheta$, $z=r\cos\vartheta$ the wave
     functions take the form
     \begin{equation}
     \psi\sim r^\ell P_\ell^{|\ell_z|}(\cos\vartheta)e^{i\ell_z\varphi}
     \prod_{i=m-n_3+1}^m
     \left(\frac{r^2}{a^2+\theta_i}-1\right)
     \label{eq:5.9}
     \end{equation}
     where
     \begin{equation}
     \ell=\alpha+\beta+\gamma+2n_1+2n_2
     =
     \alpha+\beta+\gamma+2(m-n_3)
     \label{eq:5.10}
     \end{equation}
     is the total angular momentum quantum number. Clearly $n_3$ now is the
     radial quantum number.
     \end{enumerate}

     Using the force equilibrium (\ref{eq:5.8}) it can be shown that
     \begin{equation}
     \prod_{i=m-n_3+1}^m\left(\frac{r^2}{a^2+\theta_i}-1\right)
      =
      (-1)^{n_3}\ {_2F}_1
     \left(
      -n_3,n_3+\ell+\frac{3}{2};\ell+\frac{3}{2},\frac{r^2}{a^2}
     \right)
     \label{eq:5.11}
     \end{equation}
     and, furthermore, that
     \begin{equation}
     \sum_{i=m-n_3+1}^m\frac{1}{\theta_i}
     = -\left(\ell+\frac{3}{2}\right)\sum_{i=m-n_3+1}^m\frac{1}{a^2+\theta_i}
     = n_3\left(n_3+\ell+\frac{3}{2}\right)\,.
     \label{eq:5.12}
     \end{equation}
     This simplifies the result (\ref{eq:5.7}) for the mode frequencies
      which        now becomes
     \begin{equation}
     \omega^2=\omega_0^2\big[2n_3^2+2n_3\ell+3n_3+\ell\big]
     \label{eq:5.13}
     \end{equation}
     and is indeed the result originally derived by Stringari \cite{String},
     of course in a much more direct way.

%%%%%%%%%%%%%%%%%%%%%%%%%%%%%%%%%%%%%%%%%%%%%%%%%%%%%
\section{Conclusion}
%%%%%%%%%%%%%%%%%%%%%%%%%%%%%%%%%%%%%%%%%%%%%%%%%%%%%
We have provided a complete solution for the eigenmodes and eigenfrequencies
of the anisotropic wave equation governing the low-frequency collisionless
density waves in Bose-Einstein condensates in harmonic oscillator traps with
arbitrary anisotropy in the Thomas-Fermi limit. The eigenfrequencies are given
by eq.(\ref{eq:4.13}), the mode functions are given by eq.(\ref{eq:4.8}),
where the parameters $\theta_i$ are the solutions of eqs.(\ref{eq:4.24}),
(\ref{eq:4.25}). The solution was possible, because the system was found to
be completely integrable, with three mutually commuting operators
$\hat{G},\hat{A},\hat{B}$. However, unlike in many more familiar examples in
quantum mechanics, the solution was not constructed by directly solving the
simultaneous eigenvalue equations for the three commuting operators, because
their eigenvalues, apart from that of $\hat{B}$, turned out to be rather
complicated, not providing us with simple quantum numbers. Rather our
solution proceeded by first constructing a simple form of the total wave
function which followed from the separation ansatz. The solution to the
equations fixing the free parameters $\theta_i$ in the wave function then
provided us with the natural simple integer quantum numbers $n_1,n_2,n_3$ of
the problem, on which the eigenvalues of $A$ and the mode frequencies depend
indirectly and in a complicated way. We never even had to determine the
eigenvalues of $A$ explicitely. The form (\ref{eq:4.8}) of the wave function
is remarkably reminiscent of the Bethe-ansatz. It might be interesting to
follow this connection further as it might shed more light on the mathematical
structure of the problem we have solved in this paper, and might e.g. allow
us to gain a deeper understanding of the degeneracy of the operator $\hat{B}$.
The method of solution, and in fact even the detailed structure of the
solution, generalizes directly to the analogous problem in an arbitrary number
of dimensions.. Thus the physical problem we have considered is found to be a
member of a whole family of integrable systems with connections, as we have
discussed, to billards on a curved space conformal to Euclidean space, and
the class of integrable systems discussed in the memoir of Moser \cite{Mo}.

\acknowledgments
This work has been supported by the project of the Hungarian
Academy of Sciences and the Deutsche Forschungsgemeinschaft under Grant
No. 95. R.~G. wishes to acknow\-ledge support by the Deutsche
Forschungsgemeinschaft through the Sonderforschungsbereich 237 ``Unordnung
und gro{\ss}e Fluktuationen''. A.~Cs. would like to acknowledge
support by the Hungarian National Scientific Research Foundation under
Grant Nos. OTKA T0256866, T017493 and F020094.

\begin{table}
\begin{tabular}{ccccccr}
$\alpha$ & $\beta$ & $\gamma$ & $n_1$ & $n_2$ & $n_3$ & $\omega/\omega_1$ \\
\hline
 0 & 0 & 0 & 0 & 0 & 0 &    .000000000 \\
 1 & 0 & 0 & 0 & 0 & 0 &   1.000000000 \\
 0 & 1 & 0 & 0 & 0 & 0 &   1.414213562 \\
 0 & 0 & 0 & 1 & 0 & 0 &   1.530733729 \\
 1 & 1 & 0 & 0 & 0 & 0 &   1.732050808 \\
 1 & 0 & 0 & 1 & 0 & 0 &   1.962737606 \\
 0 & 0 & 1 & 0 & 0 & 0 &   2.000000000 \\
 0 & 1 & 0 & 1 & 0 & 0 &   2.049183003 \\
 0 & 0 & 0 & 0 & 1 & 0 &   2.236067977 \\
 1 & 0 & 1 & 0 & 0 & 0 &   2.236067977 \\
 0 & 0 & 0 & 2 & 0 & 0 &   2.317848003 \\
 1 & 1 & 0 & 1 & 0 & 0 &   2.349243597 \\
 0 & 1 & 1 & 0 & 0 & 0 &   2.449489743 \\
 0 & 0 & 1 & 1 & 0 & 0 &   2.497427729 \\
 1 & 0 & 0 & 0 & 1 & 0 &   2.506633735 \\
 1 & 0 & 0 & 2 & 0 & 0 &   2.616731105 \\
 0 & 1 & 0 & 2 & 0 & 0 &   2.626705731 \\
 1 & 1 & 1 & 0 & 0 & 0 &   2.645751311 \\
 1 & 0 & 1 & 1 & 0 & 0 &   2.763607413 \\
 0 & 0 & 0 & 1 & 1 & 0 &   2.798371663 \\
\end{tabular}
\caption{Mode frequencies for the trap  
$\omega_1^2:\omega_2^2:\omega_3^2=1:2:4$ in units of $\omega_1$.  
$\alpha,\beta,\gamma$ are the parity quantum numbers, $n_1,n_2,n_3$ are the  
three positive integer quantum numbers which label uniquely each mode function  
(see text). An accidental degeneracy occurs for the states $(0,0,0,0,1,0)$  
and $(1,0,1,0,0,0)$ where $\omega/\omega_1=\sqrt{5}$.}
\end{table}
\end{document}